Corresponding Author: DOCTOR BIJAY KUMAR SHARMA, Ph.D

Corresponding Author's Institution: NATIONAL INSTITUTE OF TECHNOLOGY

First Author: BIJAY KUMAR SHARMA, Ph.D

Order of Authors: BIJAY KUMAR SHARMA, Ph.D

Abstract: In this paper I present a new perspective of the birth and evolution of Planetary Systems. This new perspective presents an all encompassing and self consistent Paradigm of the birth and evolution of the solar systems. In doing so it redefines astronomy and rewrites astronomical principles. Kepler and Newton defined a stable and non-evolving elliptical orbits. While this perspective defines a collapsing or expanding spiral orbit of planets except for Brown Dwarfs. Brown Dwarfs are significant fraction of the central star. Hence they rapidly evolve from non-Keplerian state to the end point which is a Keplerian state where it is in stable elliptical orbits. On the basis of the Lunar Laser Ranging Data released by NASA on the Silver Jubilee Celebration of Man's Landing on Moon on 21st July 1969-1994, theoretical formulation of Earth-Moon tidal interaction was carried out and Planetary Satellite Dynamics was established. It was found that this mathematical analysis could as well be applied to Star and Planets system and since every star could potentially contain an extra-solar system, hence we have a large ensemble of exoplanets to test our new perspective on the birth and evolution of solar systems. Till date 403 exoplanets have been discovered in 390 extra-solar systems by radial velocity method, by transiting planet method, by gravitational lensing method, by direct imaging method and by timing method. I have taken 12 single planet systems , 4 Brown Dwarf - Star systems and 2 Brown Dwarf pairs. Following architectural design rules are corroborated through this study of exoplanets. All planets are born at inner Clarke's Orbit what we refer to as inner geo-synchronous orbit in case of Earth-Moon System. The inner Clarke's Orbit is an orbit of unstable equilibrium. By any perturbative force such as cosmic particles or radiation pressure, the planet gets tipped long of aG1 or short of aG1. Here aG1 is inner Clarke's Orbit. If planet is long of aG1 then it is said to be in extra-synchronous orbit. Here Gravitational Sling Shot effect is in play. In gravity assist planet fly-by maneuver in space flights, gravitational sling shot is routinely used to boost the space craft to its destination. The exoplanet can either be launched on death spiral as CLOSE HOT JUPITERS or can be launched on an expanding spiral path as the planets in our Solar System are. In death spiral, exoplanet less than 5 mJ will get pulverized and vaporized in close proximity to the host star. If the mass is between 5 mJ and 7.5 mJ then it will be partially vaporized and partially engulfed by the host star and if it is greater than 7.5 mJ , then it will be completely ingested by the host star. In the process the planet will deposit all its material and angular momentum in the Host Star. This will leave tell-tale imprints of ingestion: in such cases host Star will have higher 7Li, host star will become a rapidly rotating progenitor and the host star will have excess IR. All these have been confirmed by observations of Transiting Planets. It was also found that if the exo-planet are significant fraction of the host star then those exo-planets rapidly migrate from aG1 to aG2 and have very short Time Constant of Evolution as Brown Dwarfs have. But if exo-planets are insignificant fraction of  the host star as our terrestrial planets are then they are stay put in their original orbit of birth. By corollary this implies that Giant exoplanets reach nearly Unity Evolution Factor in a fraction of the life span of a solar system. This is particularly true for brown dwarfs orbiting main sequence stars. In this study 4 star systems hosting Brown Dwarfs, 2 Brown Dwarf pairs and 12 extrasolar systems hosting Jupiter sized planets are selected. In Brown Dwarfs evolution factor is invariably UNITY or near UNITY irrespective of their respective age and Time Constant of Evolution is very short of the order of year or tens of years. In case of 12 exoplanets system with increasing mass ratio evolution factor increases and time constant of evolution shortens from Gy to My though there are two exceptions. TW Hydrae is a special case. This Solar System is newly born system which is only 9 million years old. Hence its exoplanet has just been born and it is very near its birth place just as predicted by my hypothesis.. In fact it is only slightly greater than aG1. This vindicates our basic premise that planets are always born at inner Clarke's Orbit. This study vindicates the design rules which had been postulated at 35th COSPAR Scientific Assembly  in 2004 at Paris, France, under the title "New Perspective on the Birth & Evolution of Solar Systems".

# The Architectural Design Rules of Solar Systems based on the New Perspective

Bijay Kumar Sharma

Electronics & Communication Department
National Institute of Technology
Patna-800005, India.
e-mail : electronics@nitp.ac.in

**ABSTRACT:**

In this paper I present a new perspective of the birth and evolution of Planetary Systems. This new perspective presents an all encompassing and self consistent Paradigm of the birth and evolution of the solar systems. In doing so it redefines astronomy and rewrites astronomical principles. Kepler and Newton defined a stable and non-evolving elliptical orbits. While this perspective defines a collapsing or expanding spiral orbit of planets except for Brown Dwarfs. Brown Dwarfs are significant fraction of the central star. Hence they rapidly evolve from non-Keplerian state to the end point which is a Keplerian state where it is in stable elliptical orbits. On the basis of the Lunar Laser Ranging Data released by NASA on the Silver Jubilee Celebration of Man's Landing on Moon on 21$^{st}$ July 1969-1994, theoretical formulation of Earth-Moon tidal interaction was carried out and Planetary Satellite Dynamics was established. It was found that this mathematical analysis could as well be applied to Star and Planets system and since every star could potentially contain an extra-solar system, hence we have a large ensemble of exoplanets to test our new perspective on the birth and evolution of solar systems. Till date 403 exoplanets have been discovered in 390 extra-solar systems by radial velocity method, by transiting planet method, by gravitational lensing method, by direct imaging method and by timing method. I have taken 12 single planet systems, 4 Brown Dwarf - Star systems and 2 Brown Dwarf pairs. Following architectural design rules are corroborated through this study of exoplanets. All planets are born at inner Clarke's Orbit what we refer to as inner geo-synchronous orbit in case of Earth-Moon System. The inner Clarke's Orbit is an orbit of unstable equilibrium. By any perturbative force such as cosmic particles or radiation pressure, the planet gets tipped long of $a_{G1}$ or short of $a_{G1}$. Here $a_{G1}$ is inner Clarke's Orbit. If planet is long of $a_{G1}$ then it is said to be in extra-synchronous orbit. Here Gravitational Sling Shot effect is in play. In gravity assist planet fly-by maneuver in space flights, gravitational sling shot is routinely used to boost the space craft to its destination. The exoplanet can either be launched on death spiral as CLOSE HOT JUPITERS or can be launched on an expanding spiral path as the planets in our Solar System are. In death spiral, exoplanet less than 5 $m_J$ will get pulverized and vaporized in close proximity to the host star. If the mass is between 5 $m_J$ and 7.5 $m_J$ then it will be partially vaporized and partially engulfed by the host star and if it is greater than 7.5 $m_J$, then it will be completely ingested by the host star. In the process the planet will deposit all its material and angular momentum in the Host Star. This will leave tell-tale imprints of ingestion: in such cases host Star will have higher [7]Li, host star will become a rapidly rotating progenitor and the host star will have excess IR. All these have been confirmed by observations of Transiting Planets. It was also found that if the exo-planet are significant fraction of the host star then those exo-planets rapidly migrate from $a_{G1}$ to $a_{G2}$ and have very short Time Constant of Evolution as Brown Dwarfs have. But if exo-planets are insignificant fraction of the host star as our terrestrial planets are then they are stay put in their original orbit of birth. By corollary this implies that Giant exoplanets reach nearly Unity Evolution Factor in a fraction of the life span of a solar system. This is particularly true for brown dwarfs orbiting main sequence stars. In this study 4 star systems hosting Brown Dwarfs, 2 Brown Dwarf pairs and 12 extrasolar systems hosting Jupiter sized planets are selected. In Brown Dwarfs evolution factor is invariably UNITY or near UNITY irrespective of their respective age and Time Constant of Evolution is very short of the order of year or tens of years. In case of 12 exoplanets system with increasing mass ratio evolution factor increases and time constant of evolution shortens from Gy to My though there are two exceptions. TW Hydrae is a special case. This Solar System is newly born system which is only 9 million years old. Hence its exoplanet has just been born and it is very near its birth place just as predicted by my hypothesis.. In fact it is only slightly greater than $a_{G1}$. This vindicates our basic premise that planets are always born at inner Clarke's Orbit. This study vindicates the design rules which had been postulated at 35[th] COSPAR Scientific Assembly in 2004 at Paris, France, under the title "New Perspective on the Birth & Evolution of Solar Systems".

## 1. INTRODUCTION

From early 60's there has been a ***search for extra-terrestrial intelligence (SETI).*** This search resulted in several sightings of ***unidentified flying objects (UFO)***. There were reports of ***flying saucers*** landing on Earth. In 1947 in Roswell, New Mexico, there was a crash landing of a flying saucer. This was covered up by US State Department hence the matter was unresolved and flying saucers remained a mystery. Today the same site has been chosen as the port for Space Tourism and within five year time Virgin Galactic Spacecraft will take tourists on space travels for payments.

The debate about ***extraterrestrial intelligence*** continued and it was argued that if indeed there is ***extraterrestrial intelligence*** elsewhere there must be Earth-like planets in our Milky Way Galaxy. It was also argued that SETI must concentrate in those regions of our Galaxy where Earth-like planets are most likely to be found by anthromorphic principles. By anthromorphic principles the best places to find life in our galaxy could be on planets that orbit the Red Dwarf Star. Gliese 876 falls in this category. It is one-third the mass of our Sun and only 15 light years distant from us. It is three planet system. The planets falling in " Goldilocks Zone" around these Red Dwarfs will have maximum probability of ***extraterrestrial intelligence***. These zones are the area around the star which is neither hot nor cold for liquid water to stay . The full lifecycle of a star is dependent on its mass. The lifecycle is inversely proportional to the mass. The massive stars are short lived, their life being of million years. The light stars like Red Dwarf star are very long lived, their life cycle extend up to 100 billion years. Therefore Red Dwarf planetary system has a greatest chance of harboring an evolved form of life. Thus the idea of Extra-Solar Systems and Exo-Planets were born. Extra-Solar Systems are the Solar –Systems around other main-sequence stars and members of the extra solar –systems are exo-planets.

The following table gives the types of Stars and the likelihood of finding extra-solar systems:

**Table 1. The types of stars and the likelihood of extra-solar systems with different types.(Zimmerman 2004)**

| Types | Mass | Likelihood |
|---|---|---|
| F- Type | 1.3 to 1.5 $M_\Theta$ | 10% |
| G- Type (sun like) | 1 $M_\Theta$ | 7% |
| K-Type | 0.3 to 0.7 $M_\Theta$ | 3 to 4% |

| M-Type | 0.1 to 0.3 $M_\Theta$ | Unlikely. |
| --- | --- | --- |

M Dwarf or Red dwarf stars are most abundant outnumbering sun-like G Type stars by 10 to 1. Since these stars are likely to have earth like planets falling in Goldilocks Zone hence they are the primary target for SETI missions.

## 2. THE DISCOVERY OF FIRST EXTRA-SOLAR SYSTEM.(Lissauer 2002)

In 1986, two proposals came from the University of Arizona and the University of Perkin-Elmer for space based direct imaging of Extra-Solar Systems using 16m- infrared telescope and optical telescope respectively.(Shiga 2004, Zimmerman 2004). Atmospheric turbulence smears the star's light into an arcsecond blob and reduces the resolution therefore ground based imaging of exo-planets was impossible.

Adaptive Optics overcomes the atmospheric turbulence. Adaptive optics measures the scrambling due to air turbulence with a special sensor, then sends the information to a flexible mirror that deforms and undulates many times a second to tidy up the image. The rapid changes in the shape of the mirror exactly compensates the distorting effect of the churning atmosphere.

Recently extreme adaptive optics has been developed. It replaces hundreds of tiny pistons that reshape current flexible mirrors with thousands of smaller ones, and correct the incoming light not hundreds but thousands of times a second. This would spot a young glowing Jupiter in a much wider orbits. The road to another earth lies through another Jupiter, hence the presence of wide orbit Jupiter will mark the stars which should be closely examined first for earth like planets and then for life and intelligence.

In 1991 the first extra-solar system around a Pulsar was discovered by Alexander Wolszczan and Dale Frail. This pulsar is PSR1257+12, a rapidly rotating neutron star about 1.4$M_\Theta$ and at a distance of 2000 to 3000 light years of our Earth. In this solar-system three planets were observed. The two planets have orbital period of a few months, small eccentricities and masses a few times as large as the mass of Earth. Third planet, innermost planet, has a period of one month and the mass is that of our Moon. Pulsars are the remnants of a dead star lying between the weight of 1.4$M_\Theta$ and 3 $M_\Theta$.

In 1994, 60-inch telescope on Palomar Mountain, coupled with primitive adaptive-optics system, imaged a brown dwarf orbiting the star Gliese 229. The brown dwarf was orbiting the host star at a semi-major axis of 40AU(Astronomical Unit) where 1AU is 1.496×10$^8$ km. The same system was photographed by Hubble Space Telescope. The ground-based imaging of this binary-star was confirmed by space image. This established the technical feasibility of taking ground-based images of sub-stellar objects using telescopes fitted with adaptive-optics. A brown dwarf is a failed star and its mass lies between 13 M$_J$ to 30 M$_J$ . 13 M$_J$ is the minimum mass for initiating thermonuclear fusion through gravitational collapse. Mass below 13 M$_J$ is defined as PLANET since gravitational collapse is not enough to initiate thermonuclear fusion reaction. Mass above 30M$_J$  allows the full cycle of thermonuclear fusion letting it become a main sequence star.

In 1995 Mayor and Quiloz discovered the first exo-planet orbiting the star 51Pegasi. They used ELODIE spectrograph. In this the wobbling motion of the host star is used to detect the companion object. The wobbling motion of the host star gives rise to an effective radial velocity along the line-of-sight. Hence light coming from the host star experiences Doppler Effect. When the host star is approaching us , we record a blue shifted light and when host star is receding we record a blue shifted light. The recording of the alternate blue and red shift along the time axis gives the orbital period of the exo-planet and the magnitude of the shift gives us the mass of the host star. Since we may not be having an edge-on view of the  orbital plane and the orientation radius vector  of the orbital plane may be at an angle i, the angle of inclination of the orientation vector with respect to the line-of-sight, therefore the mass observed is MSini. We do not get the true mass  of the exo-planet unless we have an edge-on view.

In 51Pegasi extra-solar system, we have the exo-planet orbiting the host star at a semi-major axis of 4.8 million miles(0.05AU). The orbital period is 4.2 days. This exo-planet is named 51 Pegasib. The mass observed, i.e. MSini , was more massive than that of Saturn.
One of the biggest drawback of Doppler Method of detection is that only Gas Giants of the size of Jupiter and Saturn can be detected.

ELODIE spectrograph has been further improved into CORALIE echlie spectrograph mounted on the 1.2m-Euler Swiss telescope at La Silla Observatory, ESO, Chile. This has been refined and exo-planets of Uranus mass have also been detected.

In 1999, a planet around HD209458 was detected by transit method. The actual mass and the size of the planet orbiting HD 209458 has been determined by combining the transit method and Doppler shift method. The density has been inferred and it is established that HD 209458b is a gas giant primarily constituted of Hydrogen just as Jupiter and Saturn are.
In 2001 the exoplanet OGLE-TR-56b detected by transit method. A polish team using 1.3m Warsaw Telescope at the Las Campanas Observatory in Chile made this discovery. In the transit method a dip in star light is caused while the exo-planet is transiting across the host star just as we record a solar eclipse when Moon is transiting across the face of Sun on NO MOON day. In the case of OGLE-TR-56b the dip occurred for 108 minutes and repeated every 1.2 days. Using 10m Keck I telescope on Mount Kea, Hawaii, the finding was confirmed by Doppler Method in January, 2003.

Both these discoveries were too close to the host star for comfort. In the classical model there was no place for gas giants to be orbiting closer than 1 to 2 AU. *These exo-planets were called hot-jupiters and they defied the conventional wisdom.*

### 3. The Planetary Architecture proposed till now.

From the earliest time the origin of Universe and the origin of our Solar System have puzzled the Scientist. With the invention of telescope a quantitative picture of our Solar System emerged but the question of origin and its architecture remained a mystery. Some of the competing theories which are currently in vogue are hydrodynamic instability theory and core accretion theory (Santos et al 2005). In both the theories the starting point of an incipient Solar System is Solar Nebula. A Solar Nebula is set in rapid rotation state. A rapidly rotating Solar Nebula flattens like a pancake into a disc of accretion. Central part gravitationally collapses into Main Sequence(MS) Star surrounded by a circumstellar disc of gas and dust also known as a disc of accretion. Conventional theory holds that planets form after the Star is fully grown and ignited. This means Star heats up the surrounding and its solar wind pushes out the lighter part of the disc. The conventional hypothesis goes on to say that beyond the snowline (Sasselov & Lecar 2000) where the gas is relatively cooler, either through hydrodynamic instability or through core accretion Gas Giant Planets are formed. After this still beyond the snowline, Ice Giants are formed. The formation of Gas Giants and Ice Giants use up majority of Gas. The remaining part of the protoplanetary disc consists of the dust and residual gas and these mainly lie in the inner part of the Solar System. The remaining dust agglomerate and eventually gravitationally accrete into the rocky planets which are known as terrestrial planets.

Over the years through new discoveries of exoplanets and through simulation studies it was realized that core accretion theory proceeds in several steps. Beyond the snowline the ice coating around the dust is in amorphous state hence they are sticky. These tend to stick into rocky cores. These rocky cores gravitationally accrete into planetary embryos of $10M_+$ where $M_+$ is the mass of our Earth. Once the rocky cores have grown to planetary embryo size, it envelopes itself with hydrogen–helium gas through runaway gravitational accretion.

This runaway process terminates with the formation of a gap in the proto-planetary disc. As the gap fills up, this process may repeat itself until all the gas is used up and only rocky cores are left. By this time accretion disc of gas and dust is completely dissipated through photo evaporation and Roberston_Poynting Drag. So we have a definite time slot in which the solar system is born or else the opportunity is missed and the host star is left without a solar system. Once the accretion disc is dissipated, we are left with proto-planetary disc of Gas Giants, Ice Giants and numerous rocky cores. These rocky cores are colliding among themselves infrequently. These infrequent collisions may lead to complete pulverization of most of the rocky cores but consolidation of some at the cost of the pulverized ones. Thus a limited number of rocky Planets emerge. This was the probable scenario of formation and growth of our Solar System in the conventional wisdom.

But the discovery of 'hot Jupiters' and the discovery of [7]Li compelled the scientists to think in a new way. The discovery of 'hot Jupiters' made it clear that either we consider the Giant planets to be born in the outer region and subsequently allow it to migrate inward by invoking planet interacting with the accretion disc and by planet to planet gravitational scattering. Or we assume the Giant Planets to be born inside and experience strong tidal interaction which launches the hot Jupiter on a rapidly collapsing spiral orbit.

In literature migration is implied in the following manner(Santos 2005).

i.      Gravitational Scattering in a multiple body system(Marzari 2002) ,
ii.     Gravitational interaction between the gaseous and/or the planetesimals disk and the planet (Lin et al 1998, Murray et al 1998) ;
iii.    Interactions between an embedded planet and a gaseous disk (Goldreich 1980) ;

Further Astronomers and Astrophysicists have identified two migration modes(Santos 2005, Lin et al 1986, Ward 1997, Tanaka et al 2002):

Type I Migration: Planet is not able to open a gap in the proto-planetary disk;

Type II Migration: Planet is massive enough to open a gap in the proto-planetary disk

Using migratory theory Type-I and migratory theory Type-II, a perspective of the architecture of solar system has been proposed based on simulation works.

Alex N.Halliday & Bernatrd J.Wood (2009) have studied Planet's early and violent beginning which indicate that first half (i.e. planetary embryo) took less than 10My and it had reducing condition but subsequent conditions took 100My and it was oxidizing. The percentage of ordinary matter in the Initial Mass Function (IMF) decide the lightness or heaviness of accretion disc & its central star. Light mass accretion disc has dwarf stars and terrestrial planets whereas high mass accretion disc has sunlike Stars , Jovian Planets and Terrestrial. They predict that 1 to $30M_+$ are rare within 1 AU forming a desert of super-earths and neptunes. Here $M_+$ is the mass of Earth.

Daniel Apes (2009) has come up with the latest constraints on the formation dates of different celestial objects in our Solar System. At time t=0 the solar nebula is born. Just before this there was a Super-Nova Explosion in our neighbourhood. This SN caused a substantial injection of dust and of radioactive nucleids (26Al,60Fe, 41Ca, 36Cl and 53Mn) into a passing pre-solar giant molecular cloud. SN shock wave set it in rapid rotation thereby compresing the giant cloud at the poles into a pan-cake. Thus the solar –nebula was born. At this point of zero time (4.5672Gya), the first refractory materials calcium and aluminum rich inclusions (CAIs) are formed. CAIs are found in primitive meteorite. CAIs formation overlap the formation of Chondrules. Chondrules are mm-sized , once molten silicate spherules. Chondrule formation continued up to 3My from the beginning. In first $4.2\pm1.3My$ , the large sized asteroids of size bigger than 500Km are formed. Earth's formation is constrained by $^{182}Hf$-$^{182}W$ and $^{146}Sm$- $^{142}Nd$ chronometry to 30-120My. Solar Nebula with embedded stars lead to Accretion Disks. IR excess show that all young stars are surrounded by massive, optically thick accretion disk. In the classical model, in first 30My all the planets are formed. In the newer findings, gas giants are formed within 30My but terrestrial planets take up to 100My to complete their formation.

M.E.Zolansky, T.J.Teja, H.Yanes et al,2006, have discovered the presence of CAI and Chondrule Fragments and virtual absence of pristine , pre-solar grains in star dust. This demonstrates that even outer parts of the solar system abound of material that had been processed at high temperatures.

Edward W Thommes, Soko Matsumura , Fredric A Rasio,2008, through simulation studies have found that Low Disk Mass and High Viscosity resulted in no gas giants. High Disk Mass with Low Viscosity resulted in numerous Gas Giants with inward migration and acquired large eccentricities.

A.Johnson, et al,2007, S.N. Cuzzi,et al,2008, and R.P. Nelson,2005, S.Ida, et al,2008, have shown that meter barrier can be broken and large size asteroids (100Km to 1000Km) can directly be born.

In the Classical Picture: dust settles into flattened disk and collects countless planetesimals growing to 1 to 10Km diameter. In second step Planetesimals collide and form Moon to Mars size Planetary Embryos.In third step the embryos smash into one another until in inner part rocky planet cores remain and in outer part Super-earth Cores remain. Classically dust stick together at low velocities to form large fractal aggregates. Further collision compacts them into pebbles. At meter size they will again collide to smithren and by Poynting Robertson Drag they will spiral in. This creates a meter- barrier.

In paradigm-breaking work, it has been shown that planetesimals can form big , 100km to 1000km, due to self gravity of small bodies highly concentrated in the turbulent structures of proto-planetary disks. The excitation of eccentricities and inclination due to the turbulence of the disk does not affect the collisional- coagulation leading to Planetary Embryo of mass 10M+ . Sudden appearance of large planetesimals in a massive disk of small bodies boosts runaway accretion of large objects. This helps solve the problem of formation of planet embryo of the size 10M+.

Papaloizou, et al ,2007, show through simulation that the number of planets that form and amount of migration and dynamical evolution they undergo depend on the mass of disk and lifetime. Short lived, low-mass disks tend to produce few giant planets and undergo limited orbital migration. More massive disks would produce more giant planets that undergo large scale migration and appreciable dynamical interaction.

Reidemeister et al ,2009, fails to explain how HR 8799 placed the three planets in the outer region of the Solar System. They have invoked planet and protoplanetary disc interaction as well they have invoked planet and host interaction.

Through all these studies there has been no satisfactory explanation either for the hot-jupiters or widely separated Jupiter-star systems.

3. **The Planetary architecture based on Planetary-Satellite Dynamics (Sharma et al 2004)**

The author feels that no architectural perspective is complete and comprehensive until the very important ingredient of tidal interaction is included in the paradigm of Planets birth and Evolution.

Krasinky and Brumberg ,2004, have detected a secular increase in the Sun-Earth distance of 15cm/yr through the analysis of all available radiometric measurements of distances between the Earth and the major planets. This includes the observations of martian landers and orbiters over 1971-2003. This cannot be attributed to the cosmological expansion of the Universe. The General Theory of Relativity seems to have opposite effects leading to zero perturbation of the planets. The consequence of solar mass loss due to solar wind and electromagnetic radiation will lead to reduction in C (moment of inertia) and hence to only 0.3cm/yr secular increase. The only plausible explanation is tidal interaction between Sun-Planets and the need to conserve the total angular momentum leading to secular increase in semi-major axis (Takaho Miura et al ,2009).

Further evidence has emerged that Gas Giants in sub-synchronous orbits are trapped in collapsing spiral orbit through tidal interaction. These Gas Giants, having tidally induced decaying orbit, will either evaporate away on close proximity, partially evaporate away or be accreted by the host star. In case of accretion, there will be definite after signatures in the host star such as increase in the relative abundance of $^{7}$Li and the host star will become a rapid rotator. The observed cut-off in small semi-major axes(a) of the hot Jupiters is 0.1AU.

Tidal dissipation in planetary satellite dynamics was well known and a thoroughly investigated subject. In 18$^{th}$ Century, German Philosopher Kant had suggested the theory of retardation of Earth's spin based on the ancient records of Solar Eclipses (Stephenson 1986, Stephenson 2003). Similar kind of studies have been carried out by Kevin Pang at Jet propulsion Laboratory at Pasadena (Morrison 1978, Jong & Soldt 1989). He happened to step upon certain ancient records regarding Solar Eclipses. A total Solar Eclipse had been observed in the town of Anyang, In Eastern China, on June 5, 1302 B.C. during the reign of Wu Ding. Had Earth maintained the present rate of spin, the Eclipse should have been observed in middle of Europe. This implies that in 1302 B.C. i.e. 3,291 years ago Earth's spin period was shorter by 0.047 seconds. This leads to a slowdown rate of 1.428 seconds per 100,000 years.

In 1879 George Howard Darwin carried out a complete theoretical analysis of Earth-Moon System and put forward a sound hypothesis for explaining the slow down of Earth's spin on its axis (Darwin 1879, Darwin 1880).

In modern times, the latest works on the Dynamical History of E-M System have been done by George E.Williams (2000), by Sigfrido Leschiutta et al (2001) and by G.A. Krasinsky (2002).

The author has done the complete analysis of E-M system and correctly been able to give the theoretical formulation of Lengthening of Day curve as well as the spiral orbit of Moon.(Sharma et al ,2002, Sharma et al, 2009).

Jean Paul Zahn (1977) applied the tidal interaction to planet and host-star.

Carlberg et al (2009)  and Jackson et al (2007) have shown that planets in sub-synchronous orbits are caught in a collapsing spiral orbit racing to their certain doom as they spiral inward. These collapsing spiral orbits are described as Death Spirals.

For planet mass < 5mj,  orbital angular momentum transfer takes place  during orbital decay only. Planet completely evaporates in proximity of the host star.

If   5mj < Planet mass < 7.5mj   then the planet partially evaporates and partially accretes. Orbital angular momentum is transferred during orbital decay and during accretion.

If Planet mass >7.5mj, then complete accretion takes place and orbital angular momentum is completely transferred to the host. All of the star that ingest planets on Main Sequence stage  are having transiting planets. This is consistent with observation. Many transiting planets are in the midst of tidal migration. Brown Dwarf have never been observed in  death spiral. Observational signatures of accretion are: IR excess and [7]Li enrichment. It was also found that 3.5% of Red Giants would have their evolution significantly affected by planets:

1.Planets in death spiral will be ingested even before Red Giant Branch(RGB).

2. Planets in outward spiral will affect Red Giants in RGB through Horizontal Branch and through Asymptotic Giant Branch.

3. Planets may affect Red Giant by rotation induced mass-loss.

Jackson et al , 2008, have shown that large measured radius of HD209458b is due to tidal heating. GJ876d may be heated. Tidal dissipation parameter and love number defined. Patzold, et al (2008) through the study of OGLE cases have countered  the 'pile up' of close-in planets at about 0.04AU .Pile-up was put forward by Sasselov,(2003).

Carol Grady(eureka Scientific & NASA/Goddard Space Flight) have proved through simulation that isolated Stars hold on to their disk material much longer than the stars in clusters. In clusters due to irradiation , disks are lost. Protoplanet interacts tidally with the disk during its growth. When mass of the planet  grows beyond  a critical mass,  a gap is induced in the disc of accretion which checks the growth of giant.

Hellier,et al (2009)  have  predicted that inward spiral of WASP-18b will continue until it reaches Roche's Limit. At this point it will be tidally disrupted and its material will feed into the star through Lagrange Point. PHS will spin up from 1/5.6d to 1/0.7d . Planet Hosting Star  will become Rapidly Rotating Star. [7]Li abundance will increase. WASP-18b is a diagnostic planet. For Planet-Star Quality Factor= $10^6$, the transit time  should increase by 28sec in a decade. This will be tested in this decade.

Israelian, et al ,2001, have the  evidence of Planet Engulfment by the star HD82943.

**5. The Time-line of Solar System Evolution based on the most recent findings.**

**Table 2. The Timeline of Planetary Formation. [Chambers 2004].**

| Triggering | Birth of Solar Nebula | Partial Dissipation of gas & dust disk(Jupiter complete) | Last Giant Impact (Earth Formation complete) | Late Heavy Bombardment | Life |
|---|---|---|---|---|---|
| 4.568Gya | 4.567Gya | 4.558Gya | 4.468Gya | 4.0Gya | 3.568Gya |
|  | 0 | 9My | 99My | 567My | 999My |

| | |
|---|---|
| Triggering | A supernova explosion in our neighborhood generates shock waves which sets a passing-by interstellar cloud of gas and dust into spin mode. This spinning primordial cloud flattens into a pancake like disc of cloud and dust. |
| Birth of Solar Nebula<br>0 year | The dust particles may be colliding and sticking giving rise to pebble sized solids. These pebbles further coalesce to form 100 to 1000km-size planetesimals crossing the meter barrier. The formation of CAI formally marks the birth of Solar Nebula and the start of Planetary Formation. Planetesimals collide and accrete to form planetary embryos. |
| Dissipation of gas and dust disk.<br>30My | Particles less than micron size and gases are pushed out by photon pressures. This is known as Photo-evaporation. Particles of micron size and more are acted upon by Robertson-Poynting drag which constrains these particles to spiral inward and eventually fall into the host body. This leads to gradual dispersal and dissipation of gas-dust disk. By this process all the gas and dust will be removed in 30My. This means that within this narrow time slot the Gas Giants should have completed their formation. Hence in first 30My Jupiter and Saturn should complete their formation. Planetary embryos get enveloped by Hydrogen gas through gravitational accretional runaway mechanism terminated by the paucity of material because of a gaping void. When the void gets filled up then the next sequential gravitational accretional runaway process initiated. This process is repeated until all the gases are exhausted. In this way in 30My Jupiter, Saturn, Neptune and Uranus formation is completed. |
| Last Giant Impact | The terrestrial planets are not formed by runaway gravitational accretional mechanisms because such large amounts of material is not present to sustain such a process. Instead a series of infrequent and titanic impacts caused the formation of the present sized Earth, Venus, Mars and Mercury. The Giant Impact was the last such event, atleast in the context of Earth, which formally marked the completion of formation of Earth. |

| Late Heavy Bombardment Era. | About 300 My after the birth of Solar Nebula, 1:2 MMR crossing occurred by the spirally expanding orbits of Jupiter and Saturn. This caused Neptune to be flung into Kuiper Belt. The disturbance of Kuiper Belt caused a large amount of comets and asteroids to be flung into the inner part of the Solar System. This caused all the planets to experience a Late Heavy Bombardment Era at about 567My after the birth of Solar Nebula. The foot prints of this era is well preserved in the petrological record of Moon.<br>My personal communication:Theoretical formulation of the origin of cataclysmic late heavy bombardment era based on the new perspective of birth and evolution of solar systems, http://arXiv.org/abs/0807.5903, (2008) |
|---|---|
| Life | After 1 Gy the first organic life based on anaerobic fermentation was initiated. |

So the Planetary Architectural Design Rules based on Planetary Satellite Dynamics are:

1.      The Giant Planets are born sequentially, heaviest being born the first. In our Solar System Jupiter was born the earliest.

2.      Terrestrial Planets will also follow the order of descending mass.

3.      Every Planet-Star system individually has two orbits where the Planetary Orbital Period is equal to the spin period of the host star. These are called the inner Clarke's orbit($a_{G1}$) and outer Clarke's orbit ($a_{G2}$).

4.      All Planets are born at inner Clarke's orbit($a_{G1}$). The slightest perturbation causes the planet to tumble short of $a_{G1}$ or long of $a_{G1}$.

5.      If Planet is long of $a_{G1}$, it experiences a powerful impulsive torque which is akin to Gravitational Sling Shot. The rotational kinetic energy imparted by the slowing down of spinning host-star to the orbiting planet during the conservative phase helps planet to spiral out to second clarke's orbit inspite of the fact that star-planet system has become a dissipative system.

6.      At second clarke's orbit , the planet is in a stable Keplerian orbit as is the case with Pluto-Charon system or with any host star-brown dwarf system or with any brown dwarf pair.

7.      A planet can never go beyond $a_{G2}$ . If it goes, it will be immediately deflected back. This means the space beyond $a_{G2}$ is a forbidden zone for the planet.

8.      If Planet is short of $a_{G1}$, it goes in a gravitation runaway inward spiral path. The planet speeds up the central spinning star giving momentum to host star and itself losing momentum. A part of the transferred energy speeds up the central star and the remaining heats the host star. This way the planet is trapped in a death spiral doomed to photo-evaporation in close proximity to host star or to complete accretion by the host star or partial evaporation and partial ingestion.

9.     Two terms are defined Time constant of Evolution ,τ, and evolution factor,€.

10.    Time Constant of Evolution describes the rate of dynamical evolution. Where planet is insignificant fraction of the host star, the rate of evolution is on Gy time scale but where secondary body is a significant fraction of the primary as is the case in brown-dwarf pair or in star-BD system, there the rate of evolution is in years.

11.    Evolution Factor can at most be UNITY. In slowly evolving systems €<<1. But in brown-dwarf pair and in star-BD system system €~ 1.

12.    A newly born system as TW-Hydrae, we can see the planet just tumbling out of $a_{G1}$.

All the above design rules have been consistently satisfied except in a few exceptional cases.

## 6. Methodology of investigation.

We first set up the LOM/LOD equation. This terminology is taken from Earth-Moon System. In E-M system lom is the orbital period of the secondary body and lod is the spin period of the primary.

In ExoPlanet context 1/lod=ω and 1/lom=Ω . Therefore lom/lod = ω/Ω.

From Basic Mechanics of Planet Satellite interaction with special emphasis on Earth-Moon System, http://arVix.org/abs/0805.0100 :

$$\omega/\Omega = Ea^{3/2} - Fa^2 \qquad\qquad 1.$$

where a= semi-major axis of the planetary orbit;

E= $J_T/(BC)$,

$J_T$ = total angular momentum of star-planet system =

Spin angular momentum of the host star + orbital angular momentum of star-planet system + spin angular momentum of the planet =

$C\omega + 0.4m_+R_m^2\Omega^{'} + m_+/(1+m_+/M_\Theta)a^2\Omega$;

here $\Omega^{'}$ = spin angular velocity of the planet around its axis of rotation and Ω = orbital angular velocity of the Planet around the Barycenter and ω= spin angular velocity of the star around its axis of rotation and a= semi-major axis of planet's orbit and $m_+$ = mass of the exoplanet.

$R_m$ = radius of the planet.

In satellite-planet system, satellite is always in synchronous orbit or in captured rotation i.e. satellite is tidally locked with the central planet hence $\Omega' = \Omega$. But this is not the case with planet-star system.

$C = 0.4 M_\Theta R_\Theta^2$ = rotational inertia of the host star where $M_\Theta$=mass of the host star and $R_\Theta$=radius of the host star;

$B=\sqrt{(G(M_\Theta + m_+))}$ where G is the Gravitational Constant.

But generally second term in total angular momentum expression is negligible compared to the third term except in very hot Jupiter case hence

$J_T = C\omega + m_+/(1+m_+/M_\Theta)a^2\Omega$;

In very hot Jupiter case:

$J_T = C\omega + (0.4 m_+ R_m^2)\Omega' + m_+/(1+ m_+/M_\Theta )a^2\Omega$;

Where $R_m$ = radius of the planet;

And F= $m_+/(C(1+m_+/M_\Theta))$.

After the ratio equation is set up i.e. Equation 1 is set up we proceed to find the two roots of Eq.1 . Here the ratio refers to the ratio of Orbital period to stellar rotation period and this ratio equation is the central pillar of this tidal induced evolution of planet's orbit.

The two roots of

$\omega/\Omega = Ea^{3/2}-Fa^2 = 1$          2.

are $\mathbf{a_{G1}}$ and $\mathbf{a_{G2}}$ .

Kepler's Third Law namely: $\Omega^2 a^3 = G(m_+ + M_\Theta)$

is strictly applicable in these two planetary orbits only. Elsewhere planet is in a spiral path with a net radial velocity given by the equation:

$v(a) = (K/a^M)[(Ea^2 - Fa^{2.5} - \sqrt{a})(2(1+ m_+/M_\Theta)( 1/(m_+ B))$ (m/sec)

where K structure constant and M is the exponent.          3

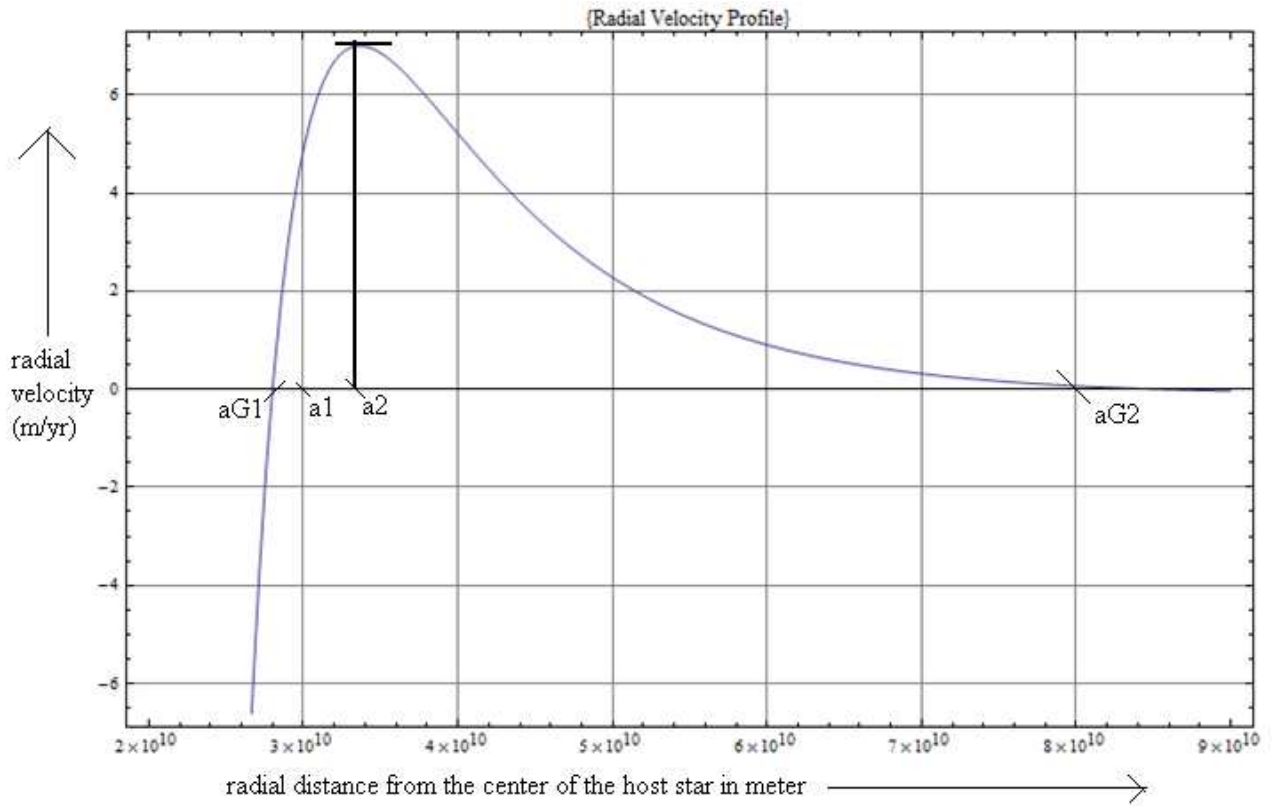

**Figure 1. Radial Velocity Profile of a Planet while it spirals out with respect to the Earth.**
**Radial Velocity Profile is given in Figure 1.**

Velocity Profile cuts the x-axis at two points. These two points are $a_{G1}$ and $a_{G2}$ or Clarke's Orbits. These are equivalent to Geo-Synchronous Orbits. In these two Orbits of semi-major axis $a_{G1}$ and $a_{G2}$, the planets and the host star are tidally locked and the spin period of the host star is equal to the orbital period of the planet and to the spin period of the planet. These two are equilibrium points but the first Clarke Orbit is unstable equilibrium and any perturbative force causes the planet to tumble short or long of the first Clarke Orbit. If the planet falls short of $a_{G1}$, planet is in sub-synchronous orbit. Planet is spinning faster than the host star and is on a gravitationally runaway collapsing spiral orbit. The planet experiences an inward radial acceleration. Finally Planet will evaporate or be swallowed by the host star.

If the planet tumbles long of $a_{G1}$, it is in super-synchronous orbit and it is spinning slower than the host star. Initially the planet-host star system is in conservative phase. At this time because of tidal interaction rotational energy and rotational angular momentum is transferred to the planet and planet experiences a rapidly increasing and outward radial acceleration. This is Gravitational Sling-Shot Phase. The positive radial acceleration is at maxima at **$a_1$**. After this maxima radial acceleration point, the angular velocity differential between the host star and the planet causes tidal friction and tidal dissipation. This leads to reduction in radial outward acceleration until it becomes zero. This is when :

$$\frac{P_2}{P1} = \frac{Orbital\ Period\ of\ Planet}{Spin\ Period\ of\ host\ star} = \frac{\omega}{\Omega} = \frac{Spin\ Angular\ Velocity\ of\ host\ star}{Orbital\ Angular\ Velocity\ of\ Planet} = 2$$

This orbital radius is symbolized as $a_2$ . This is defined as Gravitational Resonance Point. This formally marks the end of the conservative phase. At $a_2$ the outward radial velocity is at the maxima and radial acceleration is at zero. Beyond $a_2$ the radial acceleration is inward and outward radial velocity is retarded until it becomes zero at the end of non-Keplerian journey when radial velocity is zero as well as the radial acceleration is once again at zero. This is stable equilibrium state or stable Kepler state.

If the two radial velocity and radial acceleration graphs are extrapolated beyond $a_{G2}$ , both parameters are negative. That is if planet tumbles beyond $a_{G2}$ , it is immediately launched on a collapsing spiral. Thus the spatial space beyond $a_{G2}$ , is the forbidden zone just as there is forbidden zone between two permissible states of electrons orbiting the nucleus in an isolated atom.

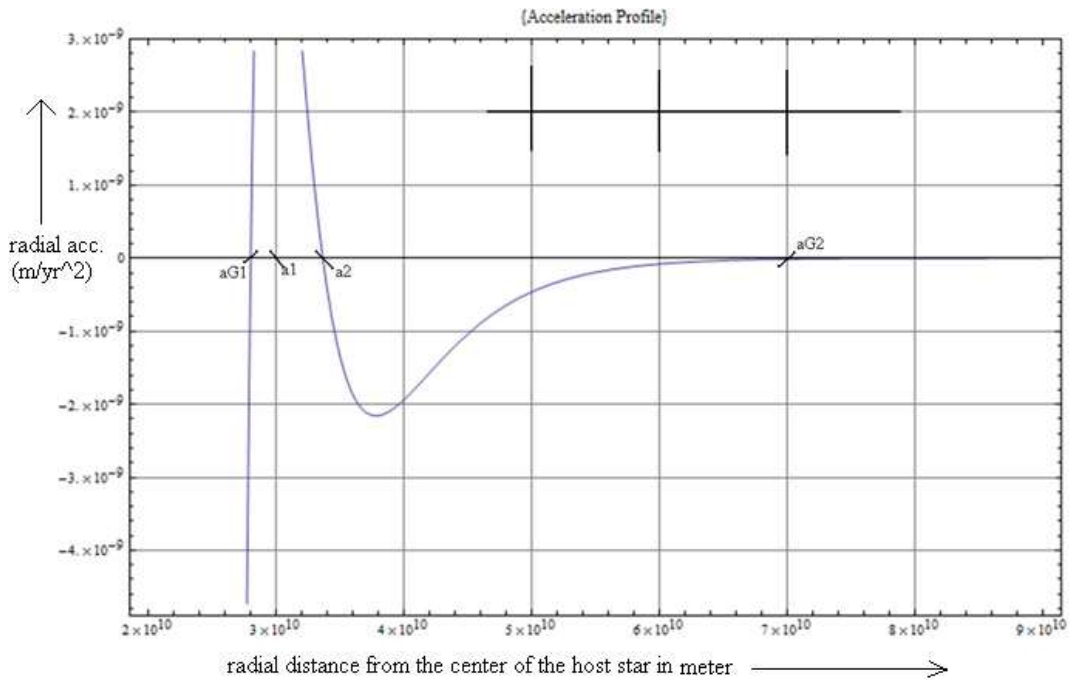

**Figure 2. Radial Acceleration Profile of a Planet while it spirals out with respect to the Earth.**

**Radial Acceleration Profile is given in Figure 2.**

Radial Acceleration Profile cuts the x-axis at three points. These three points are $a_{G1}$, $a_2$ and $a_{G2}$. At all these three points the centripetal and centrifugal are exactly balanced. Elsewhere the two forces are unbalanced. Radial Acceleration Profile has a positive maxima at **$a_1$** and a minima at some point between **$a_2$** and **$a_{G2}$**,

Kepler's Law namely, $\Omega^2 a^3 = G(M_\Theta + m_+)$, holds true only at the two Clarke Orbits. All other orbits are non-Keplerian orbits.

Keplerian orbits imply a perfect equilibrium between centripetal and centrifugal forces.

The maxima of the acceleration occurs at $a=a_1$ as shown in Figure 2 and maxima of velocity occurs at $a=a_2$ as shown in Figure 1.

Maxima of velocity is a Gravitational Resonance Point where LOM/LOD = 2 (Rubicam 1975),

i.e. $Ea^{3/2} - Fa^2 = 2$ has a root at $a = a_2$.

At $a = a_2$ the velocity of recession maxima occurs. i.e. $V(a_2) = V_{max}$.

Therefore at $a = a_2$, $(dV(a)/da)(da/dt)|_{a2} = 0$          4

Eq.(4) simplifies to the form:

$E(2-M)a^{1.5} - F(2.5-M)a^2 - (0.5-M) = 0$ at $a = a_2$          5

The root of Eq. 5 gives the exponent 'M'.

Substituting the numerical values of different constants in Eq.3, we get the algebraic expression of radial velocity of recession/approach. We determine this at $a= a_2$. This gives an algebraic expression of maximum radial velocity($V_{max}$). We arbitrarily choose a value of $V_{max}$ and solve the algebraic expression to obtain a tentative value of structure constant 'K'.

Substituting the values of exponent 'M' and structure constant 'K' in Eq 3 and multiplying by $31.557 \times 10^6$ seconds/year we obtain the expression of radial velocity of recession.

$v(a) = (K/a^M)[(Ea^2 - Fa^{2.5} - \sqrt{a})(2(1 + m_+/M_\Theta)(1/(m_+B))] \times (31.557 \times 10^6)$ (m/year)

Next we set up the time integral equation with definite integral limits between **$a_{G1}$** and present **a** namely:

$$\int_{a_{G_1}}^{a} (1 + ((K/a^M)[(Ea2 - Fa2.5 - \sqrt{a})(2(1 + m_+/M_\Theta)(1/(m_+B))] \times (31.557 \times 10^6)))) da$$

When a is greater than $a_{G1}$, planet is launched on an expanding spiral path. The above equation gives the travel time along the spiral path from $a_{G1}$ to a.

When a is shorter than $a_{G1}$, planet is launched on a contracting spiral path also known as death spiral. All 'hot Jupiters' are trapped in a similar death spiral and doomed to their destruction either by evaporation in proximity to the host star or by glancing angle collision and subsequent engulfment by the host star.

We solve the time integral equation and we get the travel time. This travel time should be of the order of the age of the host star or less than that. If the travel time exceeds the age of the host star, we go back to recalculation of structure constant 'K' by assuming a lower $V_{max}$. By several such iterations we are able to obtain a travel time less than the age of Host Star. After such an optimization we obtain the correct theoretical expression of tidal induced orbital expansion or decay. From this,

we can calculate the evolution factor $\in = \frac{a - a_{G_1}}{a_{G_2} - a_{G_1}}$   *and*

Time Constant of evolution $= \frac{a_{G_2} - a_{G_1}}{V_{max}}$ ;

### 7. The analysis of 12 Exoplanets ,4 Star-Brown Dwarf system and 2 Brown Dwarf pairs.

Below in Table 3 , 4 and5, the globe and spin parameters of the stars hosting the exo-planets, brown dwarfs  and the primary BD in case of BD pair are given.

**Table 3. The Globe and Spin Parameters of the Planet Hosting Stars(PHS).**

|    | Name of the star | Reference | Spectral type | Age(Gyr) | $P_{spin}$(d) | $M_\odot$ /$M_{sun}$ | $M_\odot$/$m_j$ | $R_\odot$ /$R_{sun}$ |
|----|------------------|-----------|---------------|----------|--------------|---------------------|-----------------|---------------------|
| 1  | HD49674          | Ref.1     | G5 V          | 2.38     | 27.2         | 1.07                | 1120.29         | 0.94                |
| 2  | HD108147         | Ref.2,3,4 | F8/G0 V       | 1.98/2.17| 8.7          | 1.27                | 1329.69         | 1.19                |
| 3  | HD52265          | Ref.5     | G0 V          | 4        | 14.6         | 1.18                | 1235.46         | 1.28                |
| 4  | HD121504         | Ref.6     | G2 V          | Geneva_1.2| 8.6         | 1.18                | 1235.46         | 1.135               |
| 5  | HD147513         | Ref.7     | G3/G5 V       | Ency_0.65| 4.7          | 0.92                | 963.58          | 1.00                |
| 6  | HD196885         | Ref.8     | F8 V          | Ency_2   | 15           | 1.33                | 1392.51         | 1.96                |
| 7  | HD196050         | Ref.9     | G3 V          | Geneva_1.6| 16          | 1.1                 | 1151.7          | 1.3                 |
| 8  | HD50554          | Ref.10    | F8            | Geneva_4.5| 16.1        | 1.11                | 1162.17         | 1.1067              |
| 9  | HD106252         | Ref.11    | G0            | Geneva_5 | 22.8         | 1.02                | 1067.94         | 1.09                |
| 10 | HD111232         | Ref.12    | G8 V          | Ency_5.2 | 30.6         | 0.78                | 816.66          | 0.95                |
| 11 | HD141937         | Ref.13    | G2/G3 V       | 2.55/1.6 | 13.25        | 1.1                 | 1151.7          | 1.0365              |
| 12 | TW Hydrae        | Ref.14    | ?             | 0.009    | 2.6          | 0.7                 | 732.9           | 0.9                 |

**Table 4.The Globe and Spin Parameters of the PHS having BrownDwarfs.**

| Name of the star | Reference | Age(Gyr) | Pspin (d) | $M_\odot$ /Msun | $M_\odot$ /$m_J$ | $R_\odot$ /Rsun |
|---|---|---|---|---|---|---|
| GL86 | Ref.15,16,17,18 | 2 | 31 | 0.79 | 733.16 | 0.855 |
| HD3651 | Ref.19 | 5.13 | 44.5 | 0.79 | 827.13 | 0.947 |
| HD196885 | Ref.20 | 2 | 15 | 1.33 | 1392.51 | 1.296 |
| *GQ LUPI* | *Ref.21* | *1My* | *1\** | *0.7* | *733.16* | *0.8* |

*The stellar rotation period given by Broeg et al (2007) is (8.45±0.2)d but it doesnot render the system mathematically tenable. To make it mathematically tractable I have taken 1 day as the stellar rotation period but this is not according to my model hence I have italiced it as an anomaly.

**Table 5. The Globe and Spin Parameters of the Primary in Brown-Dwarf pairs.**

| Name of the BD Pair | Ref. | Age | Pspin(d) | $M_\odot$/Msun | $M_\odot$ /$m_J$ | $R_\odot$ /Rsun |
|---|---|---|---|---|---|---|
| 2M1207 | Ref.22 | 8My | 0.2 | 0.025 | 26.18 | 0.09 |
| 250535-05 | Ref23 | 1My | 3.3 | 0.0572 | 59.91 | 0.675 |

| Ref.Number | PHS | Title, Author, Journal |
|---|---|---|
| 1 | HD49674 | "Seven New Keck Planets orbiting G & K dwarfs", R.Paul Butlar, Geoffrey W. Marcy, Steven S. Vogt, Debra A. Fischer, Gregory W. Henry, Gregory Laughlin & Jason T. Wright, Astrophysical Journal, 582, 1,455-466,2003 |
| 2 | HD108147 | "The CORALIE survey of southern extra-solar planets VII: Two short-period saturnian companions of HD108147 & HD168746" F.Pepi, M.Mayor, F.Gallant, D.Naef, D.Queloz, N.C. Santos, S,Udry & M. Burnet, Astronomy & Astrophysics, 388, 632-638 (2002) |
| 3 | HD108147 | "On the Age of Exoplanet Host Stars", Carlos Saffe, Mercedes Gomes & Carolina Chavero, submitted to Astronomy & Astrophysics, 4[th] October 2005. |
| 4 | HD108147 | 12 April 05 "HD108147 has a possible second companion 587±34d",F.Pepe et al,2005. |
| 5 | HD52265 | A Jupiter-mass companion to HD52265: "The CORALIE survey for southern extrasolar planets V: 3 new extrasolar planets", D. Naef, M. Mayer, F.Pepe, D. Queloz, N.C.Santos, S. Udry, M.Burnet, Astronomy & Astrophysics, 375, 205-218, 2001. |
| 6 | HD121504 | A Jovian Planet Orbiting HD121504: "The CORALIE survey for southern extra-solar planets XII: Orbital Soltions for 16 extra-solar planets detected with CORALIE", M.Mayor, S.Udry, D.Naef, |
| | | F.Pepe, D. Queloz, N.C.Santos, M. Burnet, Astronomy & Astrophysics, 415, 391-402, 2004. |
| 7 | HD147513 | A Jupiter-mass planet around HD147513: "The CORALIE survey for southern extra-solar planets XII: Orbital Soltions for 16 extra-solar planets detected with CORALIE", M.Mayor, S.Udry, D.Naef, F.Pepe, D. Queloz, N.C.Santos, M. Burnet, Astronomy & Astrophysics, 415, 391-402, 2004. |
| 8 | HD196885 | A long period eccentric planet in a multiple stellar: "The ELODIE survey for northern extra-solar planets IV:HD196885, a close Binary Star with 3.7 year planet", A.C.M. Correia, S.Udry, M.Mayor, A. Eiggenberger, D.Naef, J-L Beuyt, C. Perria, D. Queloz, J.P. Siwan, F. Pepe, N.C. Santos, D. Segransan, Astronomy & Astrophysics, 479, 271-     , 2007. |
| 9 | HD196050 | A 3 Jupiter-mass planet around HD196050: "The CORALIE survey for southern extra-solar planets XII: Orbital Soltions for 16 extra-solar planets detected with CORALIE", M.Mayor, S.Udry, D.Naef, F.Pepe, D. Queloz, N.C.Santos, M. Burnet, Astronomy & Astrophysics, 415, 391-402, 2004. |
| 10 | HD50554 | A 5.2 Jupiter-mass planet orbiting HD50554: "The ELODIE survey for northern extra-solar planets I. Six new extra-solar planet candidates", C. Perrier, J-P Sivan, D. Naef, J._L Benzit, M. Mayor, D. Queloz, S.Udry, Astronomy & Astrophysics, 410, 1039-1049, 200? |
| 11 | HD106252 | A 7.6 Jupiter-mass planet orbiting HD106252: "The ELODIE survey for northern extra-solar planets I. Six new extra-solar planet candidates", C. Perrier, J-P Sivan, D. Naef, J._L Benzit, M. Mayor, D. Queloz, S.Udry, Astronomy & Astrophysics, 410, 1039-1049, 200? |
| 12 | HD111232 | HD111232b:a planet around a velocity star. "The CORALIE survey for southern extra-solar planets XII: Orbital Soltions for 16 extra-solar planets detected with CORALIE", M.Mayor, S.Udry, D.Naef, F.Pepe, D. Queloz, N.C.Santos, M. Burnet, Astronomy & Astrophysics, 415, 391-402, 2004. |
| 13 | HD141937 | A 9.7 Jupiter-mass planet orbiting HD141937. " The CORALIE survey for southern extra-solar planets VIII. The very low-mass companions of HD141937, HD162020,HD168443,HD202206:Brown Dwarfs or 'superplanets'," S.Udry, M. Mayor, D. Naef, F.Pepe, D.Queloz, N.C.Santos, M.Brunet, Astronomy & Astrophysics, 390, 267-279, 2002. |
| 14 | TW Hydrae | "A young massive planet in a star-disk system", T. Setiawan, Th. Henning, R. Lawnhardt, A. Muller, P.Weise & M. Kurster, Letter, Nature, 451, 38-41, 3 January 2008. |

| 15 | GL86 | "4Jupiter mass planet around nearby GL86-The CORAILLE survey for southern extra-solar Planets I, a planet orbiting GL86", D. Queloz et. al. Astronomy & Astophysics, 354, 99-102, February 2000; |
| 16 | GL86 | "New Constraints on GL86" A-M Lagrange, H.Beust, S.Udry, G.Chavin & M.Meyer, Astronomy & Astrophysics, 459,955-963, 2006. |
| 17 | GL86 | "Probing long period companions to planetary host", G.Chavin et al, Astronomy & Astrophysics, No.7, 2007. |
| 18 | GL86 | "GL86B: a white dwarf orbits an exoplanet host star", M. Mugrauer & R. Neuhausen, Monthly Notices of Royal Astronomical Society (Letters), 361, Issue 1, L15-L19, published online 4 June 2005. |
| 19 | HD3651 | "HD3651B,the first directly imaged brown dwarf companion of an exoplanet host star", M.Mugrauer, A. Seifahrt, R. Newhauser & T.Mazeh, Monthly Notices of the Astronomical Society:Letters,373, 1, L31-L35. Pulished online 29[th] September 2006 |
| 20 | HD196885 | "A long period eccentric planet in a multiple star system- the ELODIE survey for Northern Extra-Solar Planets V_HD196885, a close binary star with a 3.7 year planet", A.C.M. Correia, S.Udrey et al, Astronomy & Astrophysics, 479, 271-275. |
| 21 | GQ LUPI | "Homogeneous comparison of directly imaged planet candidates: GQ Lupi, 2M1207,Aβ Pictoris", Ralph Neuhaser, from the book Multiple Stars across HR Diagram edited by G. Chauvin, A.-M. Lagrange, M.Bonavita et al. Book series: ESO Astrophysics Sympopsia, Publisher Springer Berlin/Heidelberg,2008.<br>Available at arXiv:astro-ph/0509906v |
| 22 | 2M1207 | ibid |
| 23 | 2M0535-05 | "A Surprising Reversal of Temperatures in the Brown-Dwarf Eclipsing Binaries 2MASS J 05352184-0546085", Kelvan G. Stassum, Robert D.Mathieu and Jeff A.Valenti, Astrophysical Journal, Vol. 664, Issue 2, pp. 1154-1166. |

Below in Table 6, 7 and 8 the globe, orbital and Ratio equation($\omega/\Omega$) parameters of the 12 exoplanets ,4 Brown Dwarfs and 2 Secondaries of BD pair are given.

**Table 6.The Globe and Orbital Parameters of the ExoPlanets.**

| Planet Name | a(AU) | $m_*$ /$m_j$ | $B=\sqrt{[G(M_\Theta+m_*)]}$ | $P_{orbital}$ (d) | $J_t$ | E | F | $R_{obs}$ | $R_{cal}$ |
|---|---|---|---|---|---|---|---|---|---|
| HD49674b | 0.06 | 0.12 | 1.19E+10 | 4.94 | 1.24E+42 | 2.86E-16 | 6.25E-22 | 1.83E-01 | 1.93E-01 |
| HD108147b | 0.104 | 0.4 | 1.30E+10 | 10.9 | 7.02E+42 | 7.79E-16 | 1.10E-21 | 1.25E+00 | 1.25E+00 |
| HD52265b | 0.5 | 1.05 | 1.25E+10 | 119.6 | 1.05E+43 | 1.12E-15 | 2.67E-21 | 8.20E+00 | 8.02E+00 |
| HD121504b | 0.33 | 1.22 | 1.25E+10 | 63.36 | 1.14E+43 | 1.56E-15 | 3.95E-21 | 7.37E+00 | 7.45E+00 |
| HD147513b | 1.26 | 1.00 | 1.1058755E+10 | 540.4 | 1.456E+43 | 3.71E-15 | 5.35E-21 | 1.15E+02 | 1.14E+02 |
| HD196885b | 2.63 | 2.96 | 1.33E+10 | 1347.77 | 5.64E+43 | 2.15E-15 | 2.85E-21 | 9.03E+01 | 9.01E+01 |
| HD196050b | 2.43 | 3.02 | 1.21E+10 | 1321 | 4.49E+43 | 5.17E-15 | 7.98E-21 | 8.26E+01 | 7.86E+01 |
| HD50554b | 2.41 | 5.16 | 1.22E+10 | 1293 | 7.37E+43 | 1.15522E-14 | 1.86194E-20 | 8.03E+01 | 8.06E+01 |
| HD106252b | 2.7 | 7.56 | 1.17E+10 | 1598 | 1.07E+44 | 1.966E-14 | 3.05E-20 | 7.035 E+01 | 6.871 E+01 |
| HD111232b | 1.97 | 6.8 | 1.02E+10 | 1143 | 7.17E+43 | 2.58E-14 | 4.72E-20 | 3.719 E+01 | 3.674 E+01 |
| HD141937b | 1.52 | 9.7 | 1.21368E+10 | 650 | 1.08E+44 | 1.95E-14 | 4.01E-20 | 4.90E+01 | 4.7E+01 |
| TW Hydraeb | 0.04 | 6.5 | 9.684E+09 | 3.476 | 1.53E+43 | 7.215E-15 | 5.60E-20 | 1.337E+00 | 1.335E+00 |

**Table 7. The Globe and Orbital Parameters of the BrownDwarfs**

| Dwarf Name | a(AU) | $m_*$ /$m_J$ | $B=\sqrt{[G(M_\Theta+m_*)]}$ | $P_{orbital}$(d) |
|---|---|---|---|---|
| HD3651B | 480 | 47.182 | 1.06E+10 | 4.20E+06 |
| HD196885B** | 17.23 | 349.63 | 1.49E+10 | 20235.714 |
| GL86B* | 21 | 467.5849 | 1.28E+10 | 31602.34 |
| *GQ Lupi* | *66* | *42* | *9.913624161E9* | *227587* |

| $J_t$ | E | F | $R_{obs}$ | $R_{cal}$ |
|---|---|---|---|---|
| 7.57E+45 | 2.630867E-12 | 3.104466E-19 | 9.4382E+04 | 9.43224E+04 |
| 1.267E+46 | 9.900707E-13 | 6.164744E-19 | 1349 | 1346.244 |
| 1.289E+46 | 4.518258E-12 | 2.549048E-18 | 1018.88 | 1019 |
| *2.362512E45* | *1.379491E-12* | *4.369187E-19* | *227587* | *204547* |

**Table 8. The Globe and Orbital Parameters of the Secondary of the BD Pairs.**

| Dwarf name | a(AU) | $m_*$ /$m_J$ | $B=\sqrt{(G(M_\Theta+m_*))}$ | $P_{orbital}$(d) |
|---|---|---|---|---|
| 2M1207 | 45 | 4.967 | 1.987352140E9 | 638818 |
| 2M0535-05 | 0.0496 | 37.7 | 3.517898853E9 | 9.779556 |

| $J_t$ | E | F | $R_{obs}$ | $R_{cal}$ |
|---|---|---|---|---|
| 4.093271E43 | 2.637704E-10 | 1.015911E-16 | 3194090 | 3163054.7 |
| 1.22756E43 | 3.472253E-13 | 4.374852E-18 | 2.9637 | 2.97 |

In Table 9, 10 and 11 the kinematic parameters of the evolving orbits of the 12 Exo-planets, 4 Star-Brown Dwarfs and 2 BD pairs are given. The maximum radial velocity of recession is given.

In case of Exo-planets which are beyond first Clarke's Orbit have a positive radial velocity hence have a radial velocity of recession. In cases where Exo-planets are short of first Clarke's Orbit are caught in a death spiral hence they are experiencing tidally induced orbital decay. These Exo-planets in contracting orbits have negative radial velocity hence they are approaching their respective stars. This is the case with HD49674b.

**Table 9. Kinematic Parameters of the ExoPlanets.**

| Planet Name | $a_{g1}$(m) | $a_{g2}$(m) | $a_2$(m) | K | M | $V_{max}$(m/yr) |
|---|---|---|---|---|---|---|
| HD49674b | 3.21E+10 | 1.92E+11 | 6.16E+10 | 7.81905E+49 | 2.31547 | 47 |
| HD108147b | 1.33E+10 | 4.98E+11 | 2.19E+10 | 4.57E+58 | 3.24E+00 | 1.50E+01 |
| HD52265b | 1.13E+10 | 1.72E+11 | 1.92E+10 | 9.35E+56 | 3.01E+00 | 4.00E+01 |
| HD121504b | 8.92E+09 | 1.53E+11 | 1.51E+10 | 3.2107E+57 | 3.04754 | 81.5 |
| HD147513b | 4.46E+09 | 4.802E+11 | 7.23E+09 | 5.264E+61 | 3.36029 | 10000 |
| HD196885b | 6.47E+09 | 5.69E+11 | 1.05E+10 | 6.31E+62 | 3.34E+00 | 1.725E+04 |
| HD196050b | 3.57E+09 | 4.19E+11 | 5.77E+09 | 7.82E+62 | 3.37E+00 | 7.35E+04 |
| HD50554b | 2.058E+09 | 3.85E+11 | 3.3147E+09 | 6.66345E+62 | 3.39771 | 91300 |
| HD106252b | 1.43E+09 | 4.15E+11 | 2.30E+09 | 1.614E+63 | 3.4197 | 285500 |
| HD111232b | 1.19E+09 | 2.999E+11 | 1.92E+09 | 4.18587E+62 | 3.41E+00 | 183350 |
| HD141937b | 1.45E+09 | 2.39E+11 | 2.34E+09 | 4.20E+62 | 3.39E+00 | 1.00E+05 |
| TW Hydraeb | 4.30E+09 | 1.39E+10 | 5.8368E+09 | 6.49304E+65 | 3.76E+00 | 1500 |

**Table 10. Kinematic Parameters of the Star-BrownDwarfs system.**

| Brown Dwarf | $ag_1$(m) | $ag_2$(m) | $a_2$(m) | M | K | $V_{max}$ (m/yr) |
|---|---|---|---|---|---|---|
| GL86B | 3.66728E+07 | 3.14185E+12 | 5.8249E+07 | 3.49568E+00 | 3.16249E+66 | 1.50E+11 |
| HD196885B | 1.0109E+08 | 2.5793E+12 | 1.60646E+08 | 3.49205E+00 | 6.6954E+66 | 1.50E+10 |
| HD3651B | 5.25E+07 | 7.20E+13 | 8.34E+07 | 3.50E+00 | 6.77E+68 | 8.50E+13 |
| *GQ Lupi* | *8.8085E7* | *9.96865E12* | *1.28405E8* | *3.4964* | *1.61922E70* | *6.9E14* |

**Table 11. Kinematic Parameters of the BrownDwarfs Pairs.**

| Brown Dwarf | $ag_1$(m) | $ag_2$(m) | $a_2$(m) | M | K | $V_{max}$ (m/yr) |
|---|---|---|---|---|---|---|
| 2M1207 | 2.43233E6 | 6.74126E12 | 3.86149E6 | 3.49924 | 1.4218E65 | 1E16 |
| 2M0535-05 | 2.33434E8 | 6.22571E9 | 3.88708E8 | 3.16949 | 6.25627E58 | 3E5 |

In Table 12, 13 and 14, the 12 Exo-planets , 4 Brown Dwarfs  and 2 BD Pairs are arranged in ascending order of mass ratios. The evolution factors and time-constant of evolution are given along side

**Table 12. The Evolution Factors and Time Constant of Evolution of the ExoPlanets.(The ExoPlanets are arranged in ascending order of mass ratio)**

| name | Age(G) | $m_*/M_\odot$ | Transit of exo-planet* | Time taken for formation of Exo-planet¶ | ε | τ(Gyrs) | $V_{max}$(m/yr) |
|---|---|---|---|---|---|---|---|
| HD49674b | (Ency)2.38 | 1.07E-04 | 2.355Gy | 25My | -0.144311 | 3.4Gy | 47 |
| *HD108147b* | *(Ency)1.98* | *3.01E-04* | *213My* | *1767My§* | *4.26E-03* | *2.666Gy* | *2.0E02* |
| HD52265b | (Geneva)4 | 8.50E-04 | 3.99723Gy | 2.77My | 3.94E-01 | 2.42Gy | 4E01 |
| HD121504b | (Geneva)1.2 | 9.87E-04 | 1.19082Gy | 9.18My | 2.80E-01 | 1.76Gy | 81.5 |
| HD147513b | (Ency)0.65 | 0.963E-03 | 641.87My | 8.126My | 3.868E-01 | 47.56My | 26500 |
| HD196885b | (Ency)2 | 2.13E-03 | 1.99322Gy | 6.18My | 6.89E-01 | 32.6My | 17250 |
| HD196050b | (Geneva)1.6 | 2.62E-03 | 1.596Gy | 4My | 8.66E-01 | 5.65My | 73500 |
| HD50554b | (Geneva)4.5 | 4.44E-03 | 4.49656Gy | 3.344My | 0.936938 | 4.19My | 91300 |
| HD106252b | 5.02/5 | 7.08E-03 | 4.99716Gy | 2.84MY | 0.973247 | 1.448My | 285500 |
| HD111232b | 5.12/5.2 | 8.33E-03 | 5.19512Gy | 4.88My | 0.9827 | 1.63My | 183350 |
| *HD141937b* | *2.55/1.6* | *8.42E-03* | *1.37Gy* | *230Myff* | *9.57E-01* | *1.312M* | *1.80E+05* |
| TW Hydraeb | 0.009 | 8.8657E-03 | 5.15My | 3.85Myfi | 1.75E-01 | 6.4My | 1500 |

*For every exoplanet I am assuming that fully formed exoplanet is being born at $a_{G1}$ and it immediately tumbles into a sub-synchronous orbit or in super-synchronous orbit. In either case it is launched on a collapsing spiral orbit or on expanding spiral orbit. The upper limit of the transit time is the age of planet hosting **star**.

¶I am assuming that age of planet hosting **star** is really the age of the stellar nebula. Age of the  exoplanet gives us the time before the present when fully formed exoplanet at its respective $a_{G1}$ starts its non-keplerian journey either on an contracting spiral path or on an expanding spiral path. Therefore (Age of PHS-Age of the Exoplanet) = time taken by the exoplanet to accrete to its full size after the birth of  solar nebula. I am assuming that in first 30 Myrs all giant planets are born sequentially and in 100Mys since the birth of solar nebula all the terrestrial planets are fully formed not necessarily sequentially.

§HD108147b is in highly elliptic orbit. It is likely that it has a long period companion. The companion has orbital period of 587±34 days (Pepe et al 2005). It seems first the companion was born about 1.975Gy Before the Present. About 1.767Gy after the birth of solar nebula, HD108147b is born and formed. Therefore it has a transit time of 213My. But this is a very unlikely scenario hence I consider this exoplanet not satisfying my proposed paradigm.

ff HD141937b also does not seem to fit our model because it takes 230My to form whereas the time slot allotted for Gas Giant formation is only 30My. Beyond this slot protoplanetary disc get dissipated either by Poynting-Robertson Drag or by Photo-evaporation.

fi TW Hydrae is born within 5.15My since the birth of Solar Nebula. Hence its transit time is only 3.85My. This is an acceptable scenario.

**Table 13. The Evolution Factors and Time Constant of Evolution of the BrownDwarfs.(The BDs are arranged in ascending order of mass ratio)**

| Brown Dwarf | Age(Gy) | $m_*/M_\odot$ | Transit of BD | Time taken for formation of BD | $\varepsilon$ | $\tau$(yrs) | Vmax (m/y) |
|---|---|---|---|---|---|---|---|
| *GQ Lupi* | *1My* | *0.057* | *981012y* | *18,988y* | *0.990465* | *5.27d* | *6.9E14* |
| HD3651B | 5.13Gy | 0.057 | 5.1297Gy | 300,000y | 0.999882 | 0.707y | 1.015E14 |
| HD196885B | 2Gy | 0.2524 | 1.99204Gy | 7.96My | 0.996 | 179.36y | 1.5E10 |
| GL86B | 2Gy | 0.784 | 2Gy | 0 | 0.99992 | 20.9y | 1.5E11 |

**Table 14. The Evolution Factors and Time Constant of Evolution of the BrownDwarf Pairs .(The BDs are arranged in ascending order of mass ratio)**

| Brown Dwarf | Age(Gy) | $m_*/M_\odot$ | Transit of BD | Time taken for formation of BD | $\varepsilon$ | $\tau$(yrs) | $V_{max}$ |
|---|---|---|---|---|---|---|---|
| 2M1207 | 8My | 0.2 | 8My | 0 year* | 0.998627 | 6 hours | 1E16 |
| 2M0535-05 | 1My | 0.62 | 1My | 0 year* | 0.97464 | 19197.3y | 3E5 |

*Since BD pairs are born due to gravitational instability therefore time taken for formation is taken as zero.

## 8. Discussion of the theoretical results of the design rules.

In this paper I have assumed that Gravitational Sling Shot is the dominant interaction present which decides the evolving configuration of the planetary systems. The central pillar of this new perspective is that Giant Planets are being born sequentially and that heaviest, in our case Jupiter, gets born at the earliest. It is always born at the inner Clarke's Orbit($a_{G1}$) . But it is in unstable equilibrium. Any perturbative force such as cosmic particles, radiation pressure or solar wind tips the planet either short of $a_{G1}$ or long of $a_{G1}$ . If the planet is tipped short of $a_{G1}$, it is launched as a 'hot Jupiter' on a gravitational runaway path of collapsing spiral which dooms the planet to destruction, either by evaporation if mass is less than $0.5m_J$ or by partial evaporation and partial engulfment if the mass is between $0.5m_J$ and $0.75m_J$ or by complete engulfment if the mass is greater than $0.75m_J$. The accretion of hot Jupiter will lead to the rapidly rotating progenitor of Red Giant with high [7]Li abundance.

If the planet is tipped long of $a_{G1}$ it experiences a powerful impulsive torque due to gravitational sling shot effect. As shown in Figure 1 and Figure 2, the maximum of radial acceleration occurs at $a_1$ , the spike of rotational torque. During the acceleration phase the radial velocity of recession rapidly builds up and becomes maximum at $a_2$ where the impulsive torque phase ends. This point of radial velocity maxima also happens to be the gravitational resonance where spin/orbital ratio is 2. After this point there is no further transfer of energy. The system enters a dissipative phase and the planet coasts on its own toward second Clarke's Orbit $a_{G2}$. Most of the planets are coasting in an expanding spiral path towards this destiny. But since the time constant of evolution is inversely related to the mass ratio of the planet and host star, hence heavier planets are coasting much more rapidly ,

have higher $V_{max}$ and have higher Evolution Factor. Infact Brown Dwarfs are 13m$_J$ and above. They are according to this study always having an Unity Evolution Factor and their Time Constant of Evolution is very short of the order of a year.

If we study the evolution factor , time constant of evolution and $V_{max}$ , we find that with ascending order of mass ratio of planet to star, with  stars of comparable ages, we  are having ascending order of  evolution factors, ascending order of $V_{max}$ and descending order of time-constant of evolution. HD108147 and HD141937 donot quite fit the bill. This may be due to discrepancy in the original globe and orbital parameters. TWHydraeb is a new born star hence it had no chance to tidally evolve hence we find it very near its birth place namely a$_{G1}$ and its evolution factor is 0.17 only. This vindicates one of the basic tenents of this hypothesis that planets are always born in the inner region of the solar system at inner Clark's Orbit a$_{G1}$.

## 9. Conclusions

The study of 12 Planet-Star Systems, 4 BD-Star Systems and 2 BD pair systems vindicate the main hypothesis of the Architectural Design Rules based on Planetary-Satellite Dynamics. We see that in 4 BD-Star system and 2 BD Pair system, full evolution has taken place irrespective of the age of the system. Hence GQ Lupi (1My),HD3651(5.13G), HD 196885(2Gy), GL86(2Gy) and BD pairs 2M1207(8My) ,2M0535-05(1My): all these have evolution factor of the order of 0.99 except for 2M0535-05 which has €= 097. With increasing mass ratio, time constant should have been of descending order. The time constants are  of the order of 5.27d,0.7y,179.3y,20.9y, 6 hrs and 19197.3 years but are not of descending order. In case of Planet-Star systems this rule is mostly obeyed except for a few exceptions.

## Future Works

Star-BD systems and BD pair systems need to be more thoroughly investigated to determine the laws governing their time constants of evolution. The transiting planets have to be more closely studied and seen how well these fit in Zahn's tidal dissipation formulation and in planetary –satellite dynamics formulation. Most important of all the whole spectrum of exoplanets need to be studied and examined by the two set of formulations.

## REFERENCES

Apes, Daniel,:2009, " Origin of Planetary System: Constraints and Challenges", *Earth, Moon and Planet* (2009), 105, 311-320.

Broeg,C., Schmitt, T.O.B., Guenther, E., et al.,:2007, "Rotational Period of GQ Lupi", *Astronomy & Astrophysics*, 468, Issue 3, June Part IV, 1039-1044, (2007).

Carlberg, Joben K.,Majenoski, Steven R.  & Arrar , Phil (2009) "The role of Planet Accretion in creating the next generation of Red Giant Rapid Rotators", *AstroPhysics Journal*, 8[th] June 2009.

Chambers, John E.:2004 'Planetary Accretion in the inner Solar System', *Earth and Planetary Science Letters*, 223, Issues3-4, July 2004, 241-252,

Cook, C.L.,2005, " Comment on 'Gravitational Slingshot,'by Dukla,J.J., Cacioppo, R., & Gangopadhyaya, A. [American Journal of Physics, 72(5), pp 619-621,(2004)]" *American Journal of Physics*, 73(4), pp 363, April, 2005.

Cuzzi, S.N.,Hosem, R.C., Shariff, K.,:2008, " Towards Planetesimals: Dense Chondrule Clumps in Protoplanetary Nebula", *Astrophysical J.* 687, 1432-1447, 2008.

Darwin, G. H. "On the precession of a viscous spheroid and on the remote history of the Earth," *Philosophical Transactions of Royal Society of London*, Vol. **170**, pp 447-530, 1879.

Darwin, G. H. "On the secular change in the elements of the orbit of a satellite revolving about a tidally distorted planet," *Philosophical Transactions of Royal Society of London,* Vol. **171**, pp 713-891, 1880.

Dukla,J.J., Cacioppo, R., & Gangopadhyaya, A.,:2004 " Gravitational slingshot", *American Journal of Physics*, 72(5), pp 619-621, May,2004.

Epstein, K.J.,2005, "Shortcut to the Slingshot Effect," *American Journal of Physics*, **73**(4), pp 362, April, 2005.

Goldreich,P.,Tremaine,S.,:1980, ' ' *Astrophysics J.,* 241, 425(1980);

Halliday, Alex N. & Wood, Bernatrd J. :2009, "Perspectives Geochemistry: How did Earth accrete?" , *Science,* Vol.325,No. 5936, pp 44-45, 3rd July 2009;

Hellier,C., Anderson, D.R., Cameron, A. Collier et al,:2009 "An orbital period of 0.94d for hot Jupiter planet WASP-18b", *Nature*, Letters, 460, 27th Aug 09, 1098-1100.

Ida,S. , Guillot, T., Morbeidelli, A.:2008, "Accretion & Destruction of Planetesimals in turbulent disks", *Astrophysical Journal,* 686, 1292-1301.  2008;

Israelian,G., Santos,N.C.,Meyer,M. & Rebala,R.,:2001 "Evidence of Planet Engulfment by the star HD82943", Nature, 411, 163-166 (2001).

Jackson, B., Banes,R. & Grenberg, R.,(2007) "Observational evidence for Tidal Destruction of Exoplanets", Accepted by Astrophysical Journal 2007,Feb 10.

Jackson,B., Greenberg,R. and Barnes,R.,:2008 "Tidal heating of extra-solar Planets", Accepted by *Astrophysical Journal,* 2008 Feb 26.

Johnson, A., Oishi, J.S., Low,M., Klam,H., Henmig,T., Yoidin,A.,:2007, "Rapid planetesimal formation in turbulent circumstellar disk", *Nature*,448, 1022-1025, 2007. ;

Jones, J.B.,:2005 "How does the slingshot effect work to change the orbit of spacecraft?" *Scientific American*, pp 116, November, 2005

Jong, T. de. & Soldt. W. H. van "The Earliest known Solar Eclipse record redated", *Nature*, Vol. **338**, pp. 238-240, 16[th] March 1989.

Krasinsky, G. A., "Dynamical History of the Earth-Moon System", *Celestial Mechanics and Dyanical Astronomy*, 84, 27-55, 2002.

Krasinsky, G.A. and Brumberg, V.A.:2004, "Secular Increase of Astronomical Unit from analysis of the Major Planet Motions and its Interpretation", *Celestial Mechanics and Dynamical Astronomy*, 90,267-288, 2004;

Leschiutta, S. & Tavella P., "Reckoning Time, Longitude and The History of the Earth's Rotation, Using the Moon" *Earth, Moon and Planets*, 85-86 : 225-236, 2001.

Lin,D.N.C., Bodenheimar,P., Richardson, D.C.,:1998, '        ' *Nature*, 380, 606(1998);

Lin.D.N.C., Papaloizou, J.,:1986, '                    ', *Astrophysics J.*, 309, 846 (1986);

Lissauer, Jack J. ,:2002, 'ExtraSolar planets' , *Nature*, Vol. 419, 26 September 2002, 355-358;

Marzari,F. & Weidenschilling, S.J. : 2002 " Eccentric ExtraSolar Planets: The Jumping Jupiter Model" , *ICARUS* 156, 570-579, (2002)

Miura, T. , Arakida, H., Kasai, M.  and Kuramata, S.:2009, " Secular Increase of the Astronomical Unit: a possible explanation in terms of  the total angular momentum conservation law", *Astronomy & Astrophysics,* manuscript no. miura-AU-v1.03, May 19, 2009.

Morrison L. V.,:1978 "Tidal Deceleration of the Earth's Rotation Deduced from Astronomical observations in the Period A.D. 1600 to the Present." pp. 22-27, *Tidal Friction,* edited by Brosche & Sudermann, Springer, 1978.

Murray,N., Hansen,B.,Holman,M.,Tremaine,S.,:1998 " Migrating Planets", *Science,* 279, 69 (1998),

Nelson, R.P.:2005, "On the orbital evolution of low mass protoplanets in turbulent , magnetized disks", *Astronomy & Astrophysics,* 443, 1067-1085, 2005;

Papaloizou, J.C.B., Nelson, R.P., Kley, W., Masset, F.S. and Artymovicz , P.:2007 "Disk-Planet interaction during Planet Formation", from the book *"Protostars and Planets V"*, edited by Bo Reipwith, David Jewitt, Klaus Keil, Published by University of Arizona Press (2007), pp.655-668.

Patzold,M., Carone, L. and Rauer, H.:2008 " Tidal ineractions of close-in extrasolar Planets: The OGLE cases", Accepted in *Astronomy & Astrophysics*, Feb 2.2008.

Reidemeister, M. , Kriov, A.V. ,Schmitt, T.O.B. et al,:2009, " A possible architecture of the planetary system HR 8799", *Astronomy & Astrophysics,* 503, 247-258,(2009).

Rubicam. D. P. "Tidal Friction & the Early History of Moon's Orbit", *Journal of Geophysical Research*, Vol. 80, No. 11, April 19, 1975, pp. 1537-1548.

Santos, Numo C., Benz W. & Mayor M.,:2005, "Extra-Solar Planets: Constraints for Planet Formation Model", *Science,* 310**,** 251-255, (2005),

Sasselov, D.D. and Lecar, M.:2000 " On the Snow Line in Dusty Protoplanetary Disks", *The Astrophysical Journal***,** Vol. **528**, part 1 (2000), pp 995-998;

Sasselov, D.D.,:2003, 'The new transiting planet OGLE-TR-56b: orbit and atmosphere', ' , *Asrophysical Journal*, 596, 1327, 2003.

Sharma, B. K. & Ishwar, B.:2002, 'Lengthening of Day curve could be experiencing chaotic fluctuations with implications for Earth-quake predictions', *World Space Congress-2002,* 10[th]- 19[th] October 2002, Houston, Texas, USA, Abstract 03078, p.3.

Sharma, B. K. & Ishwar, B.:2004,  'A New Perspective on the Birth and Evolution of our Solar System based on Planetary Satellite Dynamics', *35[th] COSPAR Scientific Assembly***,** 18-25[th] July 2004, Paris, France.

Sharma, B. K. & Ishwar, B.,:2004, 'Jupiter-like Exo-Solar Planets confirm the Migratory Theory of Planets' *Recent Trends in Celestial Mechanics-2004*, pp.225-231, BRA Bihar University, 1[st] – 3[rd] Novermber 2004, Muzaffarpur, Bihar.Publisher Elsiever.

Sharma, B.K. , Ishwar, B. and Rangesh, N.,:2009, 'Simulation Software for the spiral trajectory of our Moon', *Advances in Space Research,* 4 (2009), 460-466.

Shiga, David : 2004," Imaging Exoplanets", *Sky & Telescope,* April 2004, pp.45-52;

Stephenson, F. R. "Historical Eclipses and Earth's Rotation", *Astronomy & Geophysics,* Vol. 44, April 2003.

Stephenson, F. R. and Houldon, M. A.:1986, *"Atlas of Historical Eclipse Maps"***,** Cambridge University Press, 1986.

Tanaka,H., Takeuchi, T., Ward, W.R.,:2002 " Three Dimensional Interaction between a Planet and an Iso-Thermal Gaseous Disk, I.Corotation and Lindblad Torques and Plant Migration", *Astrophysical J.,* 565**,**1257-1274,(2002)Feb 1.

Thommes, Edward W., Matsumura, Soko , and Rasio, Fredric A.,: 2008, "Gas Disks to Gas Giants: Simulating the birth of Planetary Systems", *Science*, 321, 814-817, 8[th] August 2008.
Ward, W.:1997, '                              ', *Astrophysics J.,* 488**,** ,L211(1997);

Williams, George E., "Geological Constraints on the Precambrian Hisotry of Earth's Rotation and the Moon's Orbit", *Review of Geophysics,* 38, 37-59. 1/February 2000.

Zahn, J.P.,:1977 " Tidal Dissipation", *Astronomy &Astrophysics*, 57, 383,1977.

Zimmerman, Robert, : 2004, 'Exo-Earths', *Astronomy*, August 2004, 42-47.

Zolansky, M.E., Teja, T.J., Yanes, H. et al,:2006, 'Elemental compositions of Comet 81P/Wild2: Samples collected by Stardust', *Science,* 314, 1735 ,2006.

**Acknowledgment:** The  Author acknowledges AICTE financial support in form of 8017/RDII/MOD/DEG(222)/98-99 under MODROBS Scheme.

**Author Information:** Reprints and permissions information is available at npg.nature.com/reprintsandpermissions. The authors declare no competing financial interests. Correspondence and requests for materials should be addressed to B.K.Sharma

(: electronics@nitp.ac.in)

**Bijay K. Sharma** (Fellow of IETE-1985, Member IEEE-2007), Date of Birth: 01/07/1946, Place of Birth: Bela, District Sitamarhi, Bihar, India.B.Tech(Hons) 62-67,Electronics & Communication Department from Indian Institute of Technology,Kharagpur, India. M.S., 69-70, Electrical Engineering with specialization in Microelectronics from Stanford University, U.S.A.; Ph.D., 70-72, Electrical Engineering with specialization in Bio-Electronics from University of Maryland, U.S.A. Currently the author is working towards his D.Sc. degree in the field of Applied Mathematics, namely Celestial Mechanics, at BRA Bihar University, Muzaffarpur, Bihar, India.
He worked as Junior Research Fellow at Central Electronics Engineering  Research Institute, Pilani, Rajasthan, in the period 67-69. In this period he developed Tolansky Interferometer for measuring thin film thickness of the order of 200Å and Nichrome Resistance of 200 ohm/square sheet resistivity with and temperature coefficient of resistance 200p.p.m per degree. In 69-72, he completed his higher studies in U.S.A.. After returning to India he worked as Pool Officer in Electrical Engineering Department of Bihar College of Engineering, Patna, Bihar, India, during the period 72-73. This was sponsored by Council of Scientific and Industrial Research. From 73-80, he was engaged in grass root activism in the villages of Bihar. From 80-84, he served as a lecturer in E & ECE Department of I.I.T., Kharagpur. 84-85, he was Assistant Professor in Electronics & Electrical Engineering Depaertment at Birla Institute of Technology & Science. 85-86, he served as Assistant Professor and Head of the Department in Electronics Engineering Department of Institute of Engineering & Technology, Lucknow, U.P. From 86 to 97, he was engaged in rural construction activities at his village home in Maniari, Muzaffarpur, Bihar, India. From Dec 97 to date he has been actively pursuing R&D in the field of Microelectronics in Electronics & Communication Engineering Department in National Institute of Technology, Patna. Presently he is in the post of Assistant Professor and Head of the Department. He has been actively engaged with the study of  Planetary Satellites in his work for D.Sc. under Prof. Bhola Ishwar.
Dr. Sharma in course of his R & D work has written a series of papers  on Universal Hybrid –π model which will go a long way in developing and establishing  an accurate compact model for analog circuit and system simulation using BJT. In course of his D.Sc. Dr. Sharma has proposed a new theory of Solar System's birth and evolution. With the discovery of exo-planets and their inventory increasing every day both in number as well as in diversity, Celestial Physicists are at complete loss in terms of a coherent theory  which can consistently explain the diversity of exoplanets. Dr. Sharma's theory fills the gap. Also other people's work is corroborating the work of Dr. Sharma.